\documentclass{emulateapj}

\def\kms{km\,s$^{-1}$}
\def\teff{$T_{\rm eff}$}

\shorttitle{Short period binary CD$-$30 11223}
\shortauthors{Vennes et al.}

\begin{document}

\title{The shortest period \lowercase{sd}B plus white dwarf binary CD$-$30 11223 (GALEX~J1411$-$3053) $^{*}$}
\author{S. Vennes$^{1}$, A. Kawka$^{1}$, S. J. O'Toole$^{2}$, P. N\'emeth$^{1,3}$, D. Burton$^{4}$
}
\affil{$^1$ Astronomick\'y \'ustav, Akademie v\v{e}d \v{C}esk\'e republiky, Fri\v{c}ova 298, CZ-251 65 Ond\v{r}ejov, Czech Republic}
\affil{$^2$ Australian Astronomical Observatory, PO Box 915, 1670 North Ryde NSW, Australia}
\affil{$^3$ Pannon Observatory and Visitor Center, 8427 Bakonybel, Hungary}
\affil{$^4$ Faculty of Sciences, University of Southern Queensland, Toowoomba, QLD 4350, Australia}

\altaffiltext{*}{Based on observations made with ESO telescopes at the 
La Silla Paranal Observatory 
under programme ID 83.D-0540, 85.D-0866, 089.D-0864.}

\begin{abstract}
We report on the discovery of the shortest period binary comprising a hot subdwarf star 
(CD$-$30 11223, GALEX~J1411$-$3053) and a massive unseen companion. Photometric data
from the All Sky Automated Survey show ellipsoidal variations of the hot subdwarf primary
and spectroscopic series revealed an orbital period of 70.5 minutes.
The large velocity amplitude suggests the presence of a massive white dwarf in the system ($M_2/M_\odot \ga 0.77$) assuming a canonical mass for the hot subdwarf ($0.48\,M_\odot$), although a white dwarf mass as low as 0.75 $M_\odot$ 
is allowable by postulating a subdwarf mass as low as 0.44 $M_\odot$.
The amplitude of ellipsoidal variations and a high rotation velocity imposed a high-inclination to the system ($i\ga68^\circ$) and, possibly, observable secondary transits ($i\ga74^\circ$). At the lowest permissible inclination
and assuming a subdwarf mass of $\sim0.48\,M_\odot$, the total mass of the system reaches the Chandrasekhar mass limit at 1.35 $M_\odot$ and
would exceed it for a subdwarf mass above $0.48\,M_\odot$.
The system should be considered, like its sibling KPD~1930+2752, a candidate progenitor for a type Ia supernova.
The system should become semi-detached and initiate mass transfer within $\approx 30$ Myr. 
\end{abstract}


\section{Introduction}

A possible scenario for triggering type Ia supernovae involves a degenerate core collapse
following the merger of two white dwarfs \citep[see a recent summary in][and recent population syntheses by \citet{too2012}]{fer2012}. A search for double degenerate stars that could
progress toward such an outcome remains unfulfilled with most binaries either too lean or too wide
\citep[SPY program, see][]{nap2001}, with the possible exception of WD~2020$-$425 \citep{nap2005}.
A more promising variation on this scenario involves the interaction of a hot subdwarf star with a close
white dwarf companion. Mass transfered from a He-star onto a CO white dwarf may ignite and even lead to detonation of the carbon core. At an early stage these objects appear as AM CVn systems \citep[see][]{yun2008,sol2010}.  
Hot subdwarfs, also referred to as extreme horizontal branch (EHB) stars are core helium burning stars that formed
preferentially in the aftermath of close, post-interacting binaries \citep{han2002,han2003}.
The identification of the short-period subdwarf plus white dwarf binary KPD~1930+2752 \citep{max2000,gei2007} lent support to the idea,
and the expectation of
conservative mass transfer gave KPD~1930+2752 the character of a proper type Ia candidate progenitor.
A search for other massive subluminous companions to hot subdwarf stars has so far failed to
provide further candidates \citep[MUCHFUSS program, see][]{gei2011,gei2012a}.

In this letter, we report on the discovery of a close companion to the hot subdwarf CD$-$30~11223.  The orbital period is the shortest known to date for such systems.
The observations were carried-out as part of an intermediate-dispersion survey
of $\sim$40 hot, ultraviolet-selected subdwarf stars from the catalog of \citet{ven2011} 
and \citet{nem2012} and aimed at constraining the fraction of close, 
evolved binaries in the EHB population. \citet{kaw2010} described the orbital
properties of the first two systems discovered as part of this survey.

The hot hydrogen-rich subdwarf (sdB) CD$-$30~11223 is relatively bright ($V\approx 12$) and nearby ($d=250-400$ pc) and was first classified by
\citet{ven2011} in spectroscopic follow-up observations of 
ultraviolet sources detected in the {\it Galaxy Evolution Explorer} (GALEX).
Based on low-dispersion spectra, they originally determined
\teff$= 26500 \pm 2700$\,K, 
$\log{g}= 5.36 \pm 0.26$, and 
an abundance of helium $\log{\rm He/H}=-1.34\pm 0.24$. These parameters were refined by
\citet{nem2012} using higher-dispersion spectra:
\teff$= 30150^{+290}_{-310}$\,K,
$\log{g}= 5.72^{+0.04}_{-0.07}$,
$\log{\rm He/H}= -1.58^{+0.03}_{-0.05}$.
\citet{nem2012} noted that temperature and surface gravity measurements in low dispersion spectra were systematically offset from measurements obtained at higher-dispersion and are probably less reliable. 
Moreover, \citet{nem2012} reported the detection of nitrogen lines and 
measured an abundance
$\log{\rm N/H}= -4.40^{+0.29}_{-0.67}$. They found no evidence of carbon or oxygen in
the atmosphere of this star.
Remarkably, large radial velocity excursions between the two series of
spectra were evident and prompted further investigations. Independently, photometric variations in
a large ASAS data set confirmed our suspicions that this object is in a close, evolved binary.
The nature of the variations (reflection effect or tidal distortion) as well
as the nature of the companion may be established with joint phasing of photometric
and spectroscopic data. 

In the following sections we describe the spectroscopic and photometric data (Section 2) and their
analysis (Section 3). Section 3.1 gives the results of a frequency analysis of the photometric data
and Section 3.2 provides corroborating evidence of the short orbital period based on spectroscopic
data. We also measured the mass function of the secondary and rotation velocity of the primary and analyzed the ellipsoidal variations to constrain the nature of the secondary star (Section 3.3).
Finally, we summarize in Section 4.

\section{Observations}

The first identification spectra of the
hot subdwarf CD$-$30~11223 were obtained with the
European Southern Observatory (ESO) Faint Object Spectrograph and Camera (EFOSC2) attached
to the 3.6 m New Technology Telescope (NTT) at La Silla
Observatory on UT 2009 August 24 (ESO P83 program). We used grism no. 11 which
has 300 lines per mm and a blaze wavelength at 4000 \AA. The slit width
was set to 1 arcsec, which resulted in a spectral resolution
of $\sim$14 \AA\ and a coverage from 3700 to 7250 \AA.
We re-observed CD$-$30~11223 at the
NTT on UT 2010 September 21 (ESO P85 program). We employed grism no. 7 (600 lines per
mm) with a dispersion of 1.96 \AA\ per pixel. The slit width was set
at 1 arcsec resulting in a resolution of $\sim$8 \AA\ .
Although the 2009 and 2010 spectra only offered modest velocity accuracy (50 to 60 \kms)
the difference in measurements was close to 800 \kms\ suggesting the presence of a massive, close
companion to the hot subdwarf. Consequently, we secured a set of intermediate dispersion spectra,
first at Siding Spring Observatory (SSO) using the 2.3m telescope
(April 2012) and at ESO using the NTT (August-September 2012).

We obtained a set of three spectra on UT 2012 April 29 and 30 with the Wide Field Spectrograph
\citep[WiFeS][]{dop2007} attached to the 2.3m telescope at SSO.
We used the B3000 and R7000 gratings which provided
spectral ranges of 3200-5900 \AA\ at a resolution $R=\lambda/\Delta\lambda = 3000$, and 5300-7000 \AA\
at $R = 7000$, respectively. We used the RT560 dichroic beam-splitter to
separate the incoming light into its red and blue components.
We maximized the signal-to-noise of each observation by extracting the spectrum from the
most significant ($\la 6$) traces. Each trace was wavelength and flux
calibrated prior to co-addition. The spectra were wavelength
calibrated using NeAr arc spectra that were obtained following each
observation.
The exposure time was 600~s.
The two exposures acquired on April 29 were separated by only 45 minutes but revealed a velocity shift
close to 500 \kms.

Finally, we obtained complete orbital cycles at the NTT from UT 2012 August 30 to 
September 2 (ESO P89 program). 
We used grism no. 20 centered on H$\alpha$ and with a dispersion of 0.55 \AA\ per pixel in the spectral range from 6040 to 7140 \AA. We set the slit width to 0.7 arcsecond resulting in a 
$\sim 2$\AA\ resolution. We set the exposure times to 120 s in order
to minimize orbital smearing. We obtained 8 spectra on the night of August 31 and 6 spectra on September 1. The spectra revealed short ($\approx 70$ min) high-amplitude velocity variations
and appeared to cover several orbits.

The field near CD$-$30~11223 was also covered with the All Sky Automated Survey \citep[ASAS,][]{poj1997,poj2001} on numerous
occasions. A total of 1060 images were obtained as part of ASAS using the photometric V band. The star
CD$-$30~11223 was not immediately identified as a variable star.
We retrieved the photometric measurements using the ASAS all star catalogue page\footnote{http://www.astrouw.edu.pl/asas/?page=main}.

\section{Analysis}

\subsection{Photometric frequency analysis}

\begin{figure}
\centering
\includegraphics[width=1.00\columnwidth]{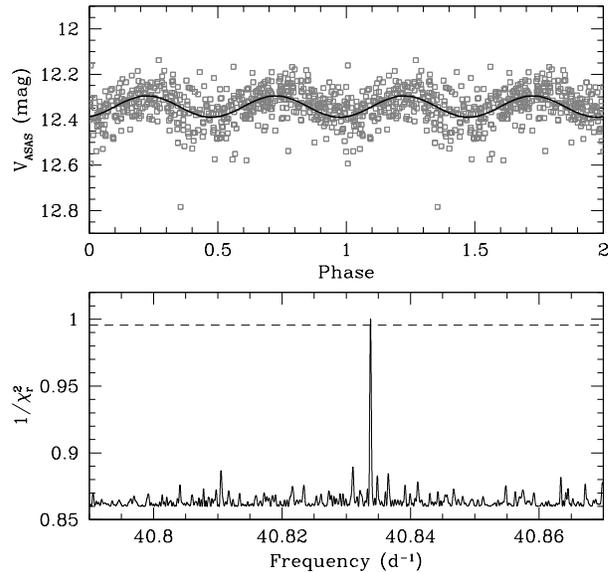}%
\caption{(Bottom panel) Frequency analysis of the ASAS photometric data (inverse of the normalized
reduced $\chi_r^2$ versus frequency) revealing
one significant frequency near 40.83 cycles\,d$^{-1}$ (full line). The 1-$\sigma$
significance level (horizontal dashed line) sets a formal statistical error on the
chosen frequency. A weaker signal (not shown) is also present at $f_{\rm phot}/2$.
(Upper panel) ASAS photometric data folded on the best frequency and best fit at $f_{\rm phot}/2$.
}
\label{fig-1}
\end{figure}

The ASAS measurements span nearly 3200 days from UT 2000 December 24 to
2009 September 13. Figure~\ref{fig-1} shows a frequency analysis of the ASAS data.
Fitting a sinusoidal function to the data set unambiguously
selected a signal frequency $f_{\rm phot}  = 40.83377 \pm 0.00003$\,d$^{-1}$ with
a semi-amplitude $\Delta m/2 = 0.047\pm0.004$ mag and mean magnitude $m_V=12.342\pm0.003$.
A reflection effect on a fainter companion would display sinusoidal
variations on the orbital period $P_{\rm orb}\ (=2\,f_{\rm phot}^{-1})$, while tidal distortion of the primary star
would induce sinusoidal variations on half the orbital period $P_{\rm orb}/2\ (=f_{\rm phot}^{-1})$ as shown in Figure~\ref{fig-1}.
An analysis of radial velocity data should help elucidate the nature of
these variations. 

\subsection{Radial velocities and secondary mass function}

All spectra were obtained at high signal-to-noise ratio and revealed well defined Balmer
and He{\sc i} lines.
We fitted the H$\alpha$ (or H$\beta$ in the P85 data) line cores using a Gaussian function.
We estimated the errors at 50-60 \kms\ in the low-dispersion data (P83 and P85) and 5-10 \kms\ in the
intermediate dispersion data (P89 and SSO). Table~\ref{tbl-1} lists all radial velocity measurements.
The heliocentric velocity is given as a function of the HJD time at mid-exposure.
Folding the data on half the photometric frequency proved successful showing that the variations
are due to tidal distortion of the bright hot subdwarf primary.
The corresponding ephemeris is
\begin{displaymath}
P=0.04897906\pm0.00000004\ {\rm d}
\end{displaymath}
\begin{equation}
T_0=1967.8426\pm0.0012
\end{equation}
where $T_0$ is the time of passage of the hot subdwarf primary at inferior conjunction.

\begin{table*}
\scriptsize
\centering
\begin{minipage}{\textwidth}
\caption{Radial velocities. \label{tbl-1}}
\centering
\begin{tabular}{ccccccccc}
\hline\hline
Telescope & HJD       & $v_{\rm sdB}$ &Telescope &  HJD   & $v_{\rm sdB}$ &Telescope & HJD   & $v_{\rm sdB}$ \\
 & ($2450000+$) & (\kms) & &  ($2450000+$) & (\kms)  &  & ($2450000+$) & (\kms) \\
\hline
ESO NTT (P83) & 5067.49158 &    375.7 &ESO NTT (P89) &   6171.46397 & $-$296.2 &ESO NTT (P89) &  6171.54961 & $-$142.9 \\
ESO NTT (P83) & 5067.49268 &    436.0 &ESO NTT (P89)&   6171.47649 &    225.6 &ESO NTT (P89) &  6172.47112 &    283.2 \\
ESO NTT (P85) & 5460.48093 & $-$377.2 &ESO NTT (P89) &   6171.48926 &    339.3 &ESO NTT (P89) &  6172.48510 & $-$329.0 \\
ESO NTT (P85) & 5460.48412 & $-$313.1 &ESO NTT (P89) &   6171.50260 & $-$245.3 &ESO NTT (P89) &  6172.49843 &  $-$61.8 \\
SSO 2.3-m & 6047.25139 & $-$340.8 &ESO NTT (P89) &   6171.51655 & $-$196.1 &ESO NTT (P89) &  6172.51202 &    415.7 \\
SSO 2.3-m & 6047.28282 &    145.6 &ESO NTT (P89) &   6171.53024 &    388.9 &ESO NTT (P89) &  6172.52588 &  $-$25.8 \\
SSO 2.3-m & 6048.01768 &    165.2 &  ESO NTT (P89) & 6171.54290 &    161.6 &ESO NTT (P89) &  6172.53945 & $-$334.2 \\
\hline
\end{tabular}\\
\end{minipage}
\end{table*}

\begin{figure}
\centering
\includegraphics[width=1.00\columnwidth]{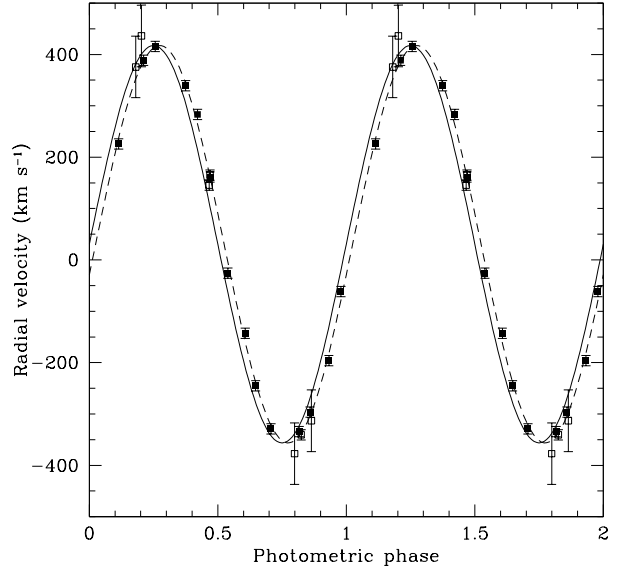}%
\caption{Radial velocity measurements (open squares: P83, P85 and SSO; full squares: P89) folded on the photometric-based ephemeris (Equation 1) compared to the predicted radial velocity curve (full line). The velocity measurements are offset from the predicted phase by only 0.0246 (dashed line) consistent with accumulated errors in the photometric ephemeris.
} 
\label{fig-2}
\end{figure}

Figure~\ref{fig-2} shows the radial velocity data folded on the (above) ephemeris.
The ASAS data preceded most of our radial velocity data by $\sim 1000$ days and the accumulated
error in the predicted phase is $\sim 1.7$ minutes. In fact, our radial velocity measurements
are offset from the predicted radial velocity curve by only 0.0246 phase (or $\sim 1.7$ minutes)
and firmly link the photometric variations to tidal distortion of the subdwarf primary.
The residual in the more precise P89 velocity measurements is 12 \kms\ and is consistent with the estimated
errors of 10 \kms.
The velocity semi-amplitude and mean velocity of the hot subdwarf are
\begin{displaymath}
\gamma_{\rm 1(sdB)} = 31.5\pm1.3\ {\rm km\,s}^{-1},
\end{displaymath}
\begin{displaymath}
K_{\rm 1(sdB)} = 386.9\pm1.9\ {\rm km\,s}^{-1},
\end{displaymath}
and the corresponding mass function for the unseen companion is
\begin{displaymath}
f_2 = 0.294\pm0.004\ M_\odot.
\end{displaymath}
Finally, we generated a phase-corrected, co-added spectrum of the H$\alpha$ and He{\sc i}$\lambda6678$
using all ESO P89 spectra. The exposure time limited orbital smearing to $\la$50 \kms, or nearly half of the
spectral resolution of the P89 data (90 \kms). A rotation velocity comparable or in excess of 100 \kms\ should be
readily measurable.  

\subsection{Properties of the primary and nature of the secondary}

\begin{figure}
\centering
\includegraphics[width=1.00\columnwidth]{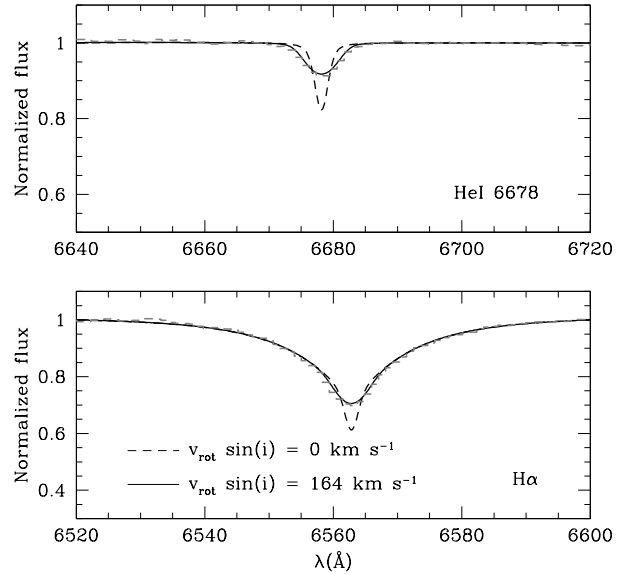}%
\caption{Co-added, intermediate dispersion spectra (grey lines) and best-fitting, rotationally broadened 
synthetic spectra (full lines) near (bottom) H$\alpha$ and (top) He{\sc i}$\lambda$6678. 
The synthetic spectra without rotation broadening are shown with dashed lines.
The effect of a projected rotation velocity of 164 \kms\ is evident in the
line cores.
} 
\label{fig-3}
\end{figure}

Approximating the full amplitude of the ellipsoidal variations by \citep[see][]{war1995}:
\begin{displaymath}
\Delta m = 3.5\, q\Big{(}\frac{R_1}{a}\Big{)}^3\, \sin^2{i}
\end{displaymath}
where $q=M_1/M_2$, $M_1$ and $R_1$ are the mass and radius of the hot subdwarf primary, and $M_2$ is the mass
of the unseen companion.
The subdwarf mass is constrained independently
but $M_2$ and $i$ are treated as free parameters.
The semi-major axis $a$ follows Kepler's third law. 
The gravity darkening and limb darkening coefficients were held fixed at values
suggested by \citet{war1995}, e.g., $\mu_g=1$ and $\mu=0.36$.

The subdwarf radius may be obtained from $R_1/R_\odot = \sqrt{M_1/(g/g_\odot)}$. Adopting $\log{g}$ from \citet{nem2012} and assuming $M_1=0.48\,M_\odot$, we estimated the subdwarf radius $R_1 = 0.159_{-0.07}^{+0.13}\,R_\odot$.
We may also set an independent and more reliable constraint on the primary radius using a measurement of the rotation velocity (Fig~\ref{fig-3}). 
Fitting synthetic H$\alpha$ profiles to co-added P89 spectra and adopting best-fit model parameters \citep{nem2012} we measured $v_{rot} \sin{i} = 164\pm5$ \kms. 
The models were first convolved with a Gaussian function with a full-width half-maximum set at 2\AA. 
Also, the effect of orbital smearing was taken into account by smoothing the models with a box-car function with a full-width of 50 \kms.
\citet{gei2012} found that all subdwarfs from their sample, single or in binaries with a period longer
than 1.2~d are slow rotators ($v_{\rm rot}\sin{i}\la 10$ \kms). Clearly,
CD$-$30~11223 is exceptional.
We verified our solution by convolving with the rotation broadening function a template built using a subdwarf (GALEX~J0206+1438) similar to CD$-$30~11223 and observed with the same set-up.
We measured an identical value of $v_{rot}\sin{i} = 164\pm15$ \kms\ but with larger error bars possibly because of a slight parameter mismatch between the two stars.
Taking observed ellipsoidal variations as evidence that the subdwarf is tidally locked to the primary, the rotation and orbital velocities are related:
\begin{displaymath}
\frac{v_{\rm rot}\sin{i}}{K_1} = \frac{R_1}{a_1},
\end{displaymath}
and since $a = a_1+a_2 = a_1(1+q)$ then
\begin{displaymath}
R_1 = \frac{a}{1+q} \frac{v_{\rm rot}\sin{i}}{K_1},
\end{displaymath}
where $q = M_1/M_2 = a_2/a_1$.
The resulting primary radius is consistent with the gravitational radius ($0.159\,R_\odot$)
for all probable inclinations ($\ga 50^\circ$). 

\begin{figure}
\centering
\includegraphics[width=1.00\columnwidth]{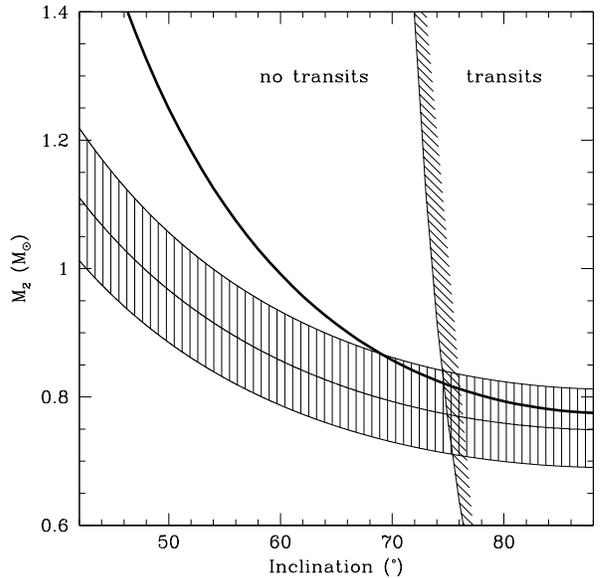}%
\caption{Constraints on the secondary mass ($M_2$) as a function of the orbital
inclination. First, $M_2$ is obtained from the measured mass function $f_2$ assuming
$M_1/M_\odot = 0.48$ (thick full line).
The range of primary mass is dictated by evolutionary 
considerations, such as the evolutionary tracks of \citet{dor1993} that
generates remnants of $0.48\,M_\odot$ \citep{zha2009,fon2012}. 
Next,
a relation between $M_2$ and the inclination is derived from the measured
amplitude of the ellipsoidal variations assuming $M_1=0.48$ (thin lines from bottom
to top
$\Delta m=$0.086, 0.094, 0.102 magnitude and shaded area).
The diagram is split between areas with detectable and non-detectable transits.
} 
\label{fig-4}
\end{figure}

Figure~\ref{fig-4} shows a set of constraints placed on the free parameters
$M_2$ and $i$. 
The mass function defines a narrow constraint for a given subdwarf mass: The minimum secondary mass ranges from 0.75 to 0.77 $M_\odot$ for a subdwarf mass
ranging from 0.44 to 0.48 $M_\odot$. 
Combining mass function and tidal constraints at $M_1=0.48\,M_\odot$, the secondary mass ranges from 0.77 to 0.87 $M_\odot$ implying a total mass of $1.25-1.35\,M_\odot$. 
The effective temperature and surface gravity
of CD$-$30~11223 place it squarely amongst the bulk of sdB stars that are best described with the models
of \citet{dor1993} within a narrow mass range centered on $\sim 0.48\,M_\odot$. 
Moreover, tests showed that constraints placed by ellipsoidal variations 
and by the mass function are incompatible below $M_1/M_\odot \approx 0.44$.
Therefore, assuming a mass close
to 0.48\,$M_\odot$ is probably justified
and the possibility that past interactions may have left an extremely low mass remnant 
\citep[$\approx0.3\,M_\odot$, see, e.g.,][]{heb2003} does not need to be considered.

Uncertainties on the rotation velocity measurements are possibly affected by
systematic effects such as over- or under-estimated spectral resolution or 
orbital smearing. Assuming a rotation velocity as low as 150 \kms\ would increase the
upper mass limit to 1.05 $M_\odot$ and push the inclination as low as 57$^\circ$.

The possibility of shallow ($\Delta m\approx (R_2/R_1)^2 \la 5$ mmag) secondary transits
cannot be ruled out and transit timing would allow us to constrain accurately the inclination and stellar radii.
Transits may become observable at an inclination $i$ such that
$\tan{(90^\circ -i)} \la R_1/a$ or $i\ga 74^\circ$ (where $a\approx 0.6\,R_\odot$).
It is worth noting that all solutions provide a primary radius
smaller than its Roche lobe radius ($R_L\approx 0.2 R_\odot$) excluding the
possibility of present-day interaction.

\section{Summary and conclusions}

We found that the hot subdwarf CD$-$30~11223 is in a close 70.5 minutes binary
with an unseen companion presumed to be a white dwarf. 
The orbital properties
and noted photometric variations suggest the presence of a companion with a
minimum mass of $0.75\,M_\odot$ assuming a minimum mass of $0.44\,M_\odot$ for the primary.  However,
the measured parameters of the sdB CD$-$30~11223 favor a canonical mass close to 0.48 $M_\odot$ that would correspond to a 
minimum
mass of 0.77 $M_\odot$ for the secondary. Constraining the radius of the primary using
a measurement of its rotation velocity and fitting the amplitude of the ellipsoidal
variations set an upper limit of 0.87 $M_\odot$ for the secondary mass at
an inclination $i\approx 68^\circ$ and a total mass
of 1.35 $M_\odot$ for the system. However, systematic effects in the
measurement of the rotation velocity and the possibility that the primary mass
may exceed 0.48 $M_\odot$ imply that the system mass may exceed the
Chandrasekhar mass limit ($M_{\rm Ch}$).
Observations of transits would help constrain the inclination of the system and
the primary radius.

Following \citet{rit1986}, the system will initiate mass transfer in
25 to 28 Myr with a contact period of 49 minutes. 
While transferring mass, the system would take the appearance of
an AM CVn system that could eventually trigger a type Ia supernova \citep{yun2008,sol2010}.
With a total life-time on the EHB that must exceed 100 Myr \citep{dor1993} and the short time remaining
before contact ($\la 30$ Myr), it is likely that a substantial amount of helium may remain in the envelope of the EHB star that would then constitute a generous helium donor.
Following \citet{nel2001}, the outcome depends on the mass of the accreting white dwarf and the
amount of helium available from the donor.
The minimum mass of the white dwarf ($\ga 0.75 \,M_\odot$) exceeds a threshold ($0.6\,M_\odot$)
below which the system would undergo nova-like, helium-flash episodes. Instead, 
should enough helium ($\ga 0.3\,M_\odot$) accumulate on the massive white dwarf after the onset
of stable mass transfer the system may suffer helium detonation that could light-up the C/O core triggering
a type Ia supernova. 
On the other hand, it is possible that the EHB star could exhaust most of its helium supply
before contact and merge with its massive white dwarf companion. Should
the total mass of the system exceed $M_{\rm Ch}$ the remnant could also trigger a type Ia
supernova. 
Alternatively, should the system mass prove insufficient, the immediate outcome would be a hot, ultramassive
white dwarf such as those sampled by \citet{ven1997} in the Extreme Ultraviolet Explorer survey.

We noted that \citet{gei2012b} independently reported on this system and reached similar conclusions. 

\acknowledgments

S.V. and A.K. acknowledge support from 
the Grant Agency of the Czech Republic (GA \v{C}R P209/10/0967, GA \v{C}R P209/12/0217). 
This work was also supported by the project RVO:67985815 in the Czech Republic.
We thank the referee for helping us clarify the evolutionary prospects of the system.
Also, we thank A. Williams (Perth Observatory) for his contribution to the project.\\

{\bf Note added in proof:} L. Yungelson informed us that, following Fink et al. (2010, A\&A, 514, 53), 
a supernova may be triggered by an accreted
mass of helium as low as $0.1\,M_\odot$ for a $0.8\,M_\odot$ white dwarf,
thereby increasing the odds of such an event taking place in CD$-$30\,11223.

\end{document}